\RequirePackage{fix-cm}

\documentclass[twocolumn]{svjour3}          

\smartqed  
\usepackage{url}
\usepackage{amssymb}
\usepackage{amsmath}
\usepackage{graphicx}
\usepackage{mathptmx}      
\usepackage{microtype}
\usepackage{latexsym}

\begin{document}

\title{Freestanding loadbearing structures with Z-shaped particles}

\author{Kieran A. Murphy, Nikolaj Reiser, Darius Choksy, Clare E. Singer, Heinrich M. Jaeger
}

\institute{K. Murphy \at
              \email{murphyka@uchicago.edu}           
           \and
}


\maketitle

\begin{abstract}
Architectural structures such as masonry walls or columns exhibit a slender verticality, in contrast to the squat, sloped forms obtained with typical unconfined granular materials. Here we demonstrate the ability to create freestanding, weight-bearing, similarly slender and vertical structures by the simple pouring of suitably shaped dry particles into a mold that is subsequently removed. Combining experiments and simulations we explore a family of particle types that can entangle through their non-convex, hooked shape. We show that Z-shaped particles produce granular aggregates which can either be fluid and pourable, or solid and rigid enough to maintain vertical interfaces and build freestanding columns of large aspect ratio ($>$10) that support compressive loads without external confinement. We investigate the stability of such columns with uniaxial compression, bending, and vibration tests and compare with other particle types including U-shaped particles and rods.  We find a pronounced anisotropy in the internal stress propagation together with strong strain-stiffening, which stabilizes rather than destabilizes the structures under load.
\end{abstract}

\section{Introduction}
\label{intro}
Dry sand is rarely the building material of choice when constructing columns or walls: the grains easily flow past each other when sheared, and while this facilitates material transport by enabling pouring and fast filling of arbitrarily shaped volumes, this fluid-like quality also results in squat, sloped piles instead of the vertical interfaces necessary for such structures.  Indeed, the free, unconfined surface of poured spherical or ellipsoidal grains at rest cannot be steeper than about 25-35 degrees, an angle controlled primarily by the degree to which friction limits slip between contacting particles \cite{Robinson2002,Train1958,Jaeger1988}.  To reach significantly larger angles of repose requires either attractive forces between particles or particle types whose shape enables `geometrical cohesion' \cite{Franklin2012} beyond frictional interactions, for example through strong interlocking or entanglement \cite{Zou2009,Reichhardt2009,Brown2012,Gravish2012,Miskin2013,Franklin2014,Jaeger2015}.  

Within the architectural context, cohesionless granular material that can support steep angles yet remain mechanically stable without external confinement opens up a host of new and intriguing possibilities. In particular, such material introduces the option to create vertical surfaces for freestanding walls and columns, or even overhangs and domed spaces, without the need for cementing or otherwise bonding together any constituent elements. This means that structures can be created by simply pouring the particles into a mold to define the overall form, and upon removing the mold the structure is freestanding and loadbearing.  Given that particle shapes that can entangle tend to produce loosely packed structures in which the interstitial void space can easily take up more than $50\%$ of the total volume, this approach combines quick deployment with low material use.  The fact that there is no bonding between particles then means the rigid structure is easily recycled -- it can be ``melted down'' and returned to a free-flowing state, ready to be reused again and again. 

These considerations depend on particle geometry and details of the filling procedure, but are independent of particle size or overall scale of the structure. Therefore, they are limited in practice mainly by how much individual particles deform under gravity and any additional load, i.e., by their internal stiffness and strength.  The central task, then, is to identify suitable particle shapes, and this is the focus of this article.

A solution is to produce jammed packings with high resistance to shear in order to mitigate failure caused by shear-induced dilation.  One way to quantify this is to consider the ratio of bulk to shear modulus, B/G, of the granular material.  Small B/G ratios correspond to small values for Poisson's ratio, which implies that compression of a material along one direction can occur with little expansion in the transverse directions. For a freestanding granular column the result is that axial (compressive) strains will lead to comparatively small radial expansion. As long as this expansion can be balanced by self-confinement to prevent particles from falling off the column, the structure will remain mechanically stable.  In fact, with increasing strain the column will jam more deeply and become mechanically stiffer, i.e., the axial compression will lead to stiffening.  Such strain-stiffening under self-confinement extends the notion of a small Poisson's ratio far beyond the regime of small deformations.

There are several approaches to try to achieve such conditions.  Here we consider only granular structures made from a single type of particle and no added components to provide tensile forces, such as string or wire wedged between particles. Particle types that satisfy this requirement include rods of sufficiently large aspect ratio \cite{Franklin2012,Philipse1996,Desmond2006,Trepanier2010}, three-dimensional many-armed stars \cite{Dierichs2012}, flexible chains (``granular polymers'' \cite{Zou2009,Reichhardt2009,Brown2012,Mohaddespour2012,Lopatina2011,Regev2013}), and rigid hook-shaped particles \cite{Gravish2012,Marschall2015}.  In all of these cases, the self-confinement after pouring, when the material has come to rest, is achieved through geometrical particle interlocking and/or entanglement.

In what follows we focus on rigid, highly non-convex particles that can entangle by hooking into each other. Prior work, in particular the detailed investigations by Dan Goldman, Scott Franklin and coworkers \cite{Gravish2012,Marschall2015}, showed that aggregates of U-shaped, ``staple''-type particles exhibit strong entanglement.  Such particles are members of a larger family of shapes that consist of a rod-like central backbone with one arm on either end (Fig. 1).  For U-shaped particles with equal-length arms perpendicular to the backbone, Gravish et al. showed that the overall aggregate stability is determined by two competing factors:  the degree of entanglement, which increases with arm length, and the packing fraction, which decreases with arm length \cite{Gravish2012}. The most stable aggregate, in the sense that it disintegrated most slowly under vibration, resulted when the arm length was chosen to be 1/2 of the length of the central backbone. 

\begin{figure}
\centering
\includegraphics[width=1.0\linewidth]{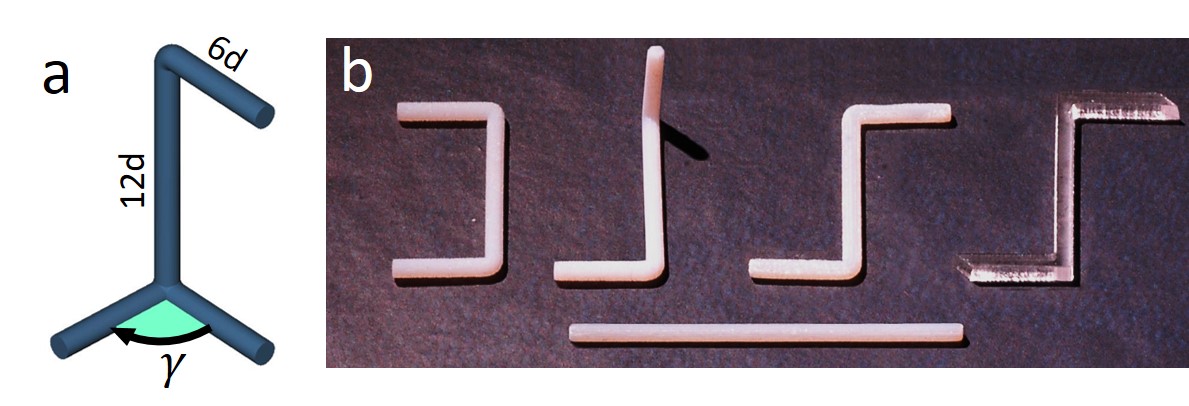}
\caption{\textbf{a} Schematic of the shapes studied, where d is the cross-sectional diameter and $\gamma$ is the angle used to define the family of shapes. \textbf{b} Particles used in the experiments: 3D printed U ($\gamma=0$), $Z_{90}$ ($\gamma=\pi/2$), Z ($\gamma=\pi$), and rod (aspect ratio 22) all with $d=1.3$mm, and an acrylic Z with $d=1.6$mm.  For a scale, the rod is 2.9cm long}
\end{figure}

\begin{figure}
\centering
\includegraphics[width=1.0\linewidth]{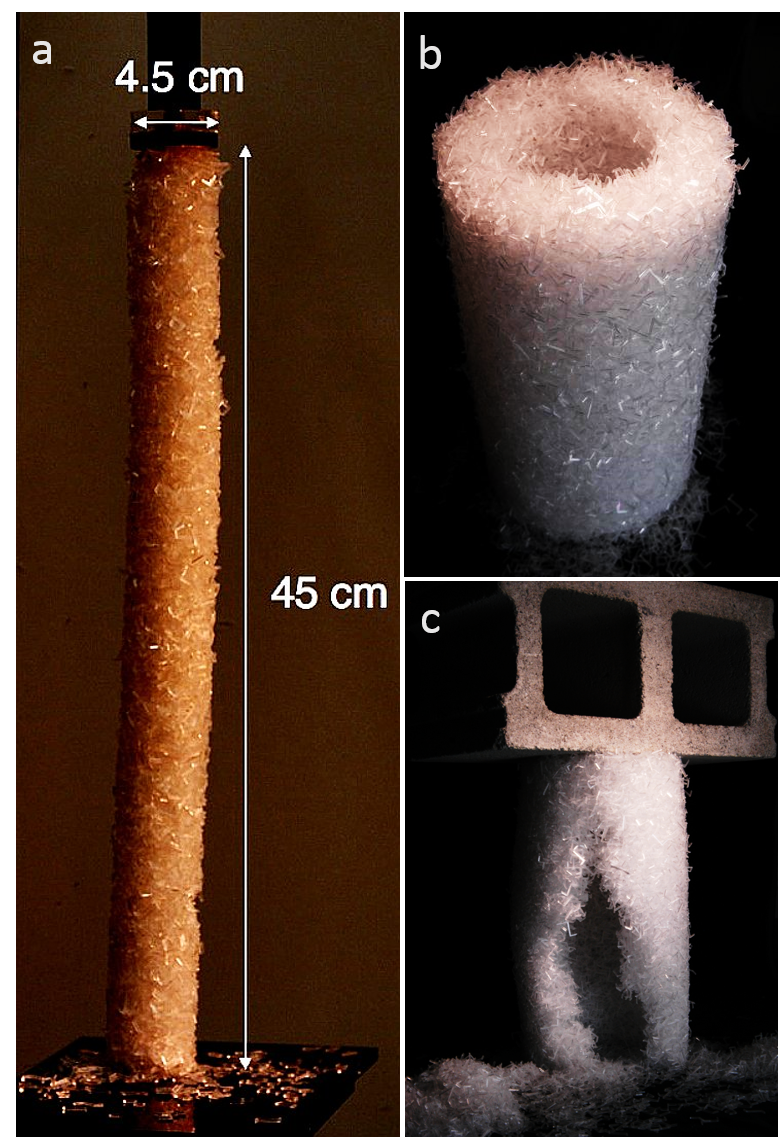}
\caption{Structures made with the acrylic Zs (d=1.6mm) \textbf{a} 10:1 freestanding column. \textbf{b} Annulus 35cm tall with 10cm inner diameter and 20cm outer diameter \textbf{c} Similar to Fig. 3b,c, we carved out a portion of the annulus to test its stability and it was still able to support a cinderblock ($\sim$13kg)}
\end{figure}

Other members of this family of shapes have one of their arms rotated out of the plane, and we investigate here how this rotation affects the aggregate stability.  Rotating by $\gamma=\pi$, one arrives at Z-shaped members of the family. As we will show, this is a particularly interesting shape that combines significant strain-stiffening and reasonable flowability in the unjammed state.  This is in contrast to particle types that entangle strongly but exhibit low flowability, and it provides an intriguing opportunity for granular, jamming-based architecture: tall, slender columns made from these particles can be constructed such that they are mechanically stable under compressive loading, yet become unstable when the load is removed. Without load, such columns therefore very easily disintegrate into their primary components, which then can be collected and reused. Last but not least, the fabrication of Z-shaped particles is made easy by their planar geometry and the fact that they tessellate 2D space, which wastes little material when cutting from large sheets.  

We compare the performance of Z-shaped particles to several other shapes, including U-shaped “staples” (rotation angle $\gamma=0$) and “twisted Z’s” ($\gamma=\pi/2$, hereafter called $Z_{90}$), in terms of their ability to form freestanding, unconfined columns that can support loads. To this end, we performed modified column collapse tests, similar to those first performed on granular materials by \cite{Lube2004} and later by \cite{Trepanier2010}.  We included axial pre-strain before removing radial confinement so as to evaluate to what degree pre-strain adds to the rigidity of the granular packing.  We also uniaxially compressed freestanding columns and found significant strain-stiffening.  Prior work using uniaxial tests on very lightly confined columns showed that for convex particle shapes as well as short-armed hexapods the stress-strain curves exhibit strain-softening \cite{Athanasi2014}; only chains which are sufficiently long exhibit enough entanglement to show strain-stiffening \cite{Brown2012}.  We complemented our experiments with molecular dynamics (MD) simulations using the granular package of LAMMPS (Large-scale Atomic/Molecular Massively Parallel Simulator)\cite{lammps}.  

Our key finding is that this family of Z-shaped particles can form solid aggregates which strongly strain-stiffen without any radial confinement and that, remarkably, past pre-strains of only a few percent these packings solidify enough to be rotated $90^{\circ}$ and can be easily formed into very slender columns with an aspect ratio (height to diameter) of 10 or more (as seen in Fig. 2).

\section{Experimental and Simulation Details}
\label{sec:1}

The family of particles tested was defined by a cylindrical backbone ($L=12d$, where $d$ is the diameter of the cross section of the particle) with a perpendicular arm at each end ($L=6d$); the angle between these arms is $\gamma$ (Fig. 1a).  We compared these particle shapes to a rod with length $L=22d$, whose volume matched that of a Z particle.  For all these particles, $d=1.3$mm (Fig. 1b).  We fabricated 2,000-3,000 particles of each shape by 3D printing them in a hard resin (Objet \textit{VeroWhite}) with Young's modulus $E_{mat}=1260 \pm 120$ MPa \cite{Athanasi2014}, using our Connex 350 Objet printer.  In addition, we laser-cut out of acrylic ($E_{mat}\sim3000$ MPa) around 30,000 Z particles with a square cross section of varying thickness ranging from $d=1.6$mm to $d=6.4$mm (Fig. 2,3). This allowed us to compare 3D printed particles with particles made from a stiffer material, and the larger quantity facilitated the construction of aggregates of many different geometries and sizes.

\begin{figure}
\centering
\includegraphics[width=1.0\linewidth]{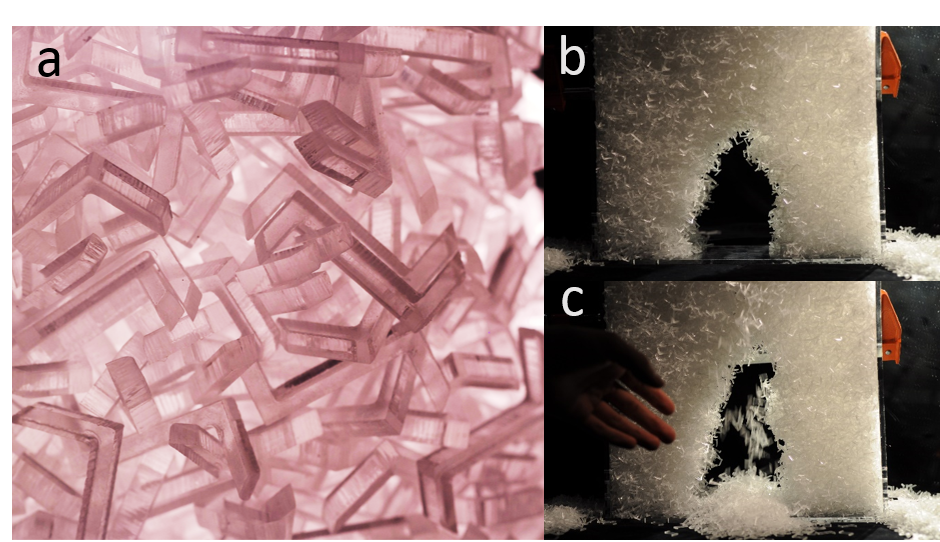}
\caption{\textbf{a} Close-up of a packing of Z particles. \textbf{b} The Z particles can indefinitely sustain an arch carved into a solid sheet with confinement only along the edges, but will flow easily when individual particles are perturbed (\textbf{c})}
\end{figure} 

All quantitative results presented in the following refer to tests on freestanding columns with an aspect ratio of at least 2:1 (height to diameter) before pre-strain was applied.  Each column was built by pouring particles into a \textit{clamshell} -- a hard plastic tube cut in half lengthwise -- and compressing the packing using an Instron 5800 Materials Tester to the desired pre-strain, $\varepsilon_0$, before removing the clamshell.  Tacks were attached to the top and bottom plates to pin the ends of the column in place.  In the uniaxial tests, $\varepsilon_0=0.03$ for every trial; strain was reset to zero upon confinement removal and compression with the Instron continued, with 5 experimental runs performed and averaged for each data point.  For the uniaxial tests, all particle shapes seen in Fig. 1b were used; other tests used only the 3D printed Z, U and $Z_{90}$ particles. 

To measure the transverse rigidity, we ran a modified 3-point bending test where the pre-strained and unconfined column was rotated $90^{\circ}$.  The Instron was then used to push quasi-statically downward at the center of the now horizontal column while it was supported at its ends. The indenter in this bending test had a radius of curvature of about 8mm.  We plot the median curve out of three experimental runs for each test point.

Finally, to quantify the stability of columns to vibration, we used a Vibration Test Systems VG-100 shaker to horizontally shake vertically oriented columns until failure, which we defined as loss of contact between the top of the column and the apparatus (which generally, but not always, was followed immediately by the column disintegrating into a pile).  As with the bending test, columns were held at fixed pre-strain for the duration of the experiment by fixing the end caps in place after pre-strain was applied.  The frequency of the shaker was fixed at 20Hz while its amplitude was ramped up such that the peak acceleration in terms of gravity, $\Gamma$, increased by 0.1 every 10 seconds until failure.  Five trials were averaged for each datapoint.

The simulations were run with particles composed of 22 rigidly bonded spheres 1.3mm in diameter and matched to experiment with 3D printed particles (including rods) in weight and size.  Interactions between the particles occurred through contacting spheres, and followed a Hertzian contact law ($F\sim\delta^{3/2}$, where $\delta$ is the overlap of the spheres) with tangential friction and collisional energy loss included.  The elastic modulus of the spheres was set to 1000MPa, the approximate stiffness of the 3D printed material, although since the spheres were rigidly bonded, there was no flexing of a particle as a whole.  About 7,000 particles were poured into a frictionless confining cylinder and then compressed via a frictional top wall to a given pre-strain before the cylinder was removed (Fig. 4).  For simulations, the confining cylinder was 10.0cm in diameter, twice as big as what was used in experiments in order to have a larger sample size for better statistics (the simulation shown in Fig. 4 was not one of these larger column sizes; it was matched to experimental dimensions).

\begin{figure}
\centering
\includegraphics[width=1.0\linewidth]{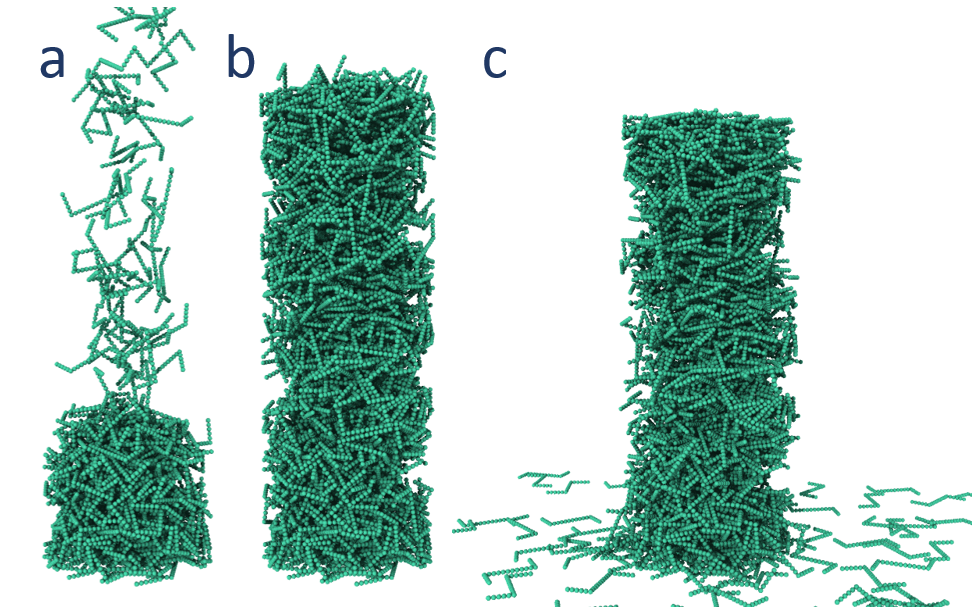}
\caption{Snapshots from a LAMMPS simulation. \textbf{a} Z particles being poured into a confining cylinder 5.0cm in diameter (not shown) \textbf{b} Completed packing after pouring and settling under gravity \textbf{c} after pre-strain $\varepsilon_0=0.1$ was applied and confinement was removed}
\end{figure}

\section{Results and Discussion}
Exploiting the two key features of Z-shaped particles -- their flowability in the unjammed state and their tendency to entangle into self-confining aggregates -- we were able to form freestanding granular structures of remarkable slenderness and stability; some of these structures are shown in Fig. 2a-c and Fig. 3b-c.  Most striking is the ability of these particles to form stable columns with large aspect ratios; Fig. 2a shows that ratios of at least 10 are possible.  Though some particles fall off as the mold is removed, their number is remarkably small, typically less than 2\% of the total.  These columns are stable under load but come apart easily by toppling and disintegration when the load is removed. This load-induced stabilization is highly unusual behavior for granular material and is in stark contrast to what is possible with any type of convex, non-entangling particle. 

\begin{figure}
\centering
\includegraphics[width=1.0\linewidth]{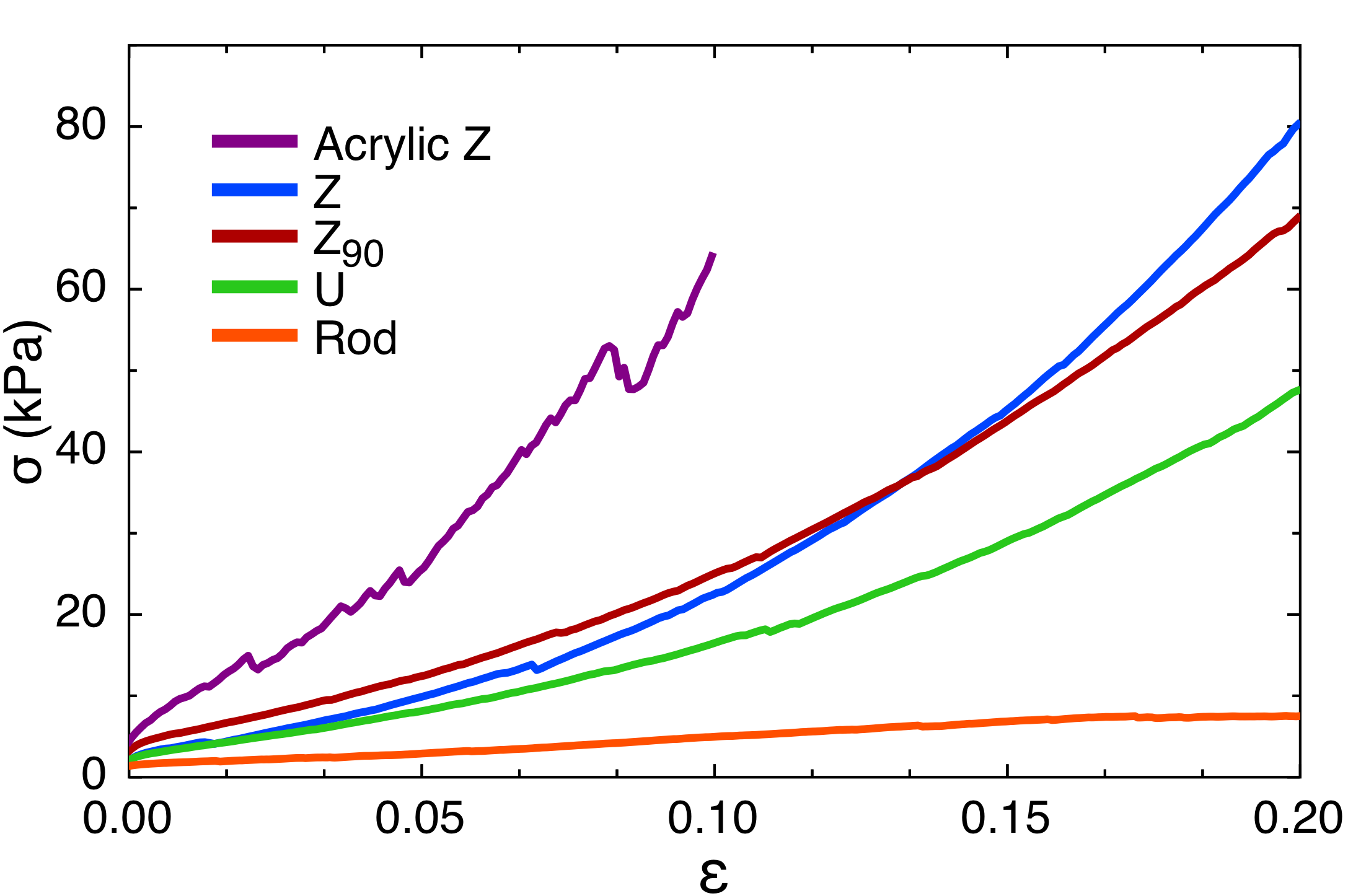}
\caption{Averaged stress-strain response under uniaxial compression for freestanding columns with pre-strain $\varepsilon_0=0.03$}
\end{figure}

It is easy to create architectural proto-structures other than columns simply by pouring the particles into spaces with different confining geometry.  Given the entangling and self-confining properties of the aggregate (which we discuss below) the confining walls feel relatively little stress from the weight of the particles and from the load applied during pre-straining.  Therefore very simple confinement, for example by thin sheets of plastic or fabric, can suffice to hold the particles in place during pouring and column-casting, making it easy to envision freeform shaping of the overall aggregate form.

With a measured packing fraction $\phi=0.26\pm0.03$ for the Zs (and slightly less for the other shapes tested: $0.22\pm0.03$ for both the $Z_{90}$ and U particles), structures are lightweight and require relatively little material to construct.  The rigidity of the granular object, being tunable by the applied pre-strain, was adjusted to fit the needs of the structure: the creation of the 10:1 column required pre-strain of several percent to hold its form without confinement, whereas an annulus (2:1 height to outer diameter, 2:1 outer diameter to inner diameter), a structure which is inherently much more stable, required no pre-strain and therefore no additional stress beyond what was provided by self-weight.  Not only was the annulus able to stand freely, but it was able to support the weight of a cinderblock ($\sim$13kg) and have a significant portion of its wall carved out before failing (Fig. 2c).  The latter aspect is particularly interesting: large openings (e.g., arches or domes) can be built-in from the start, for example by filling around a balloon which is later removed, or carved out after the structure has been cast by repeatedly agitating individual particles which then fall off and loosen the particles around them (Fig. 3c).

\begin{figure*}
\centering
\includegraphics[width=0.8\textwidth]{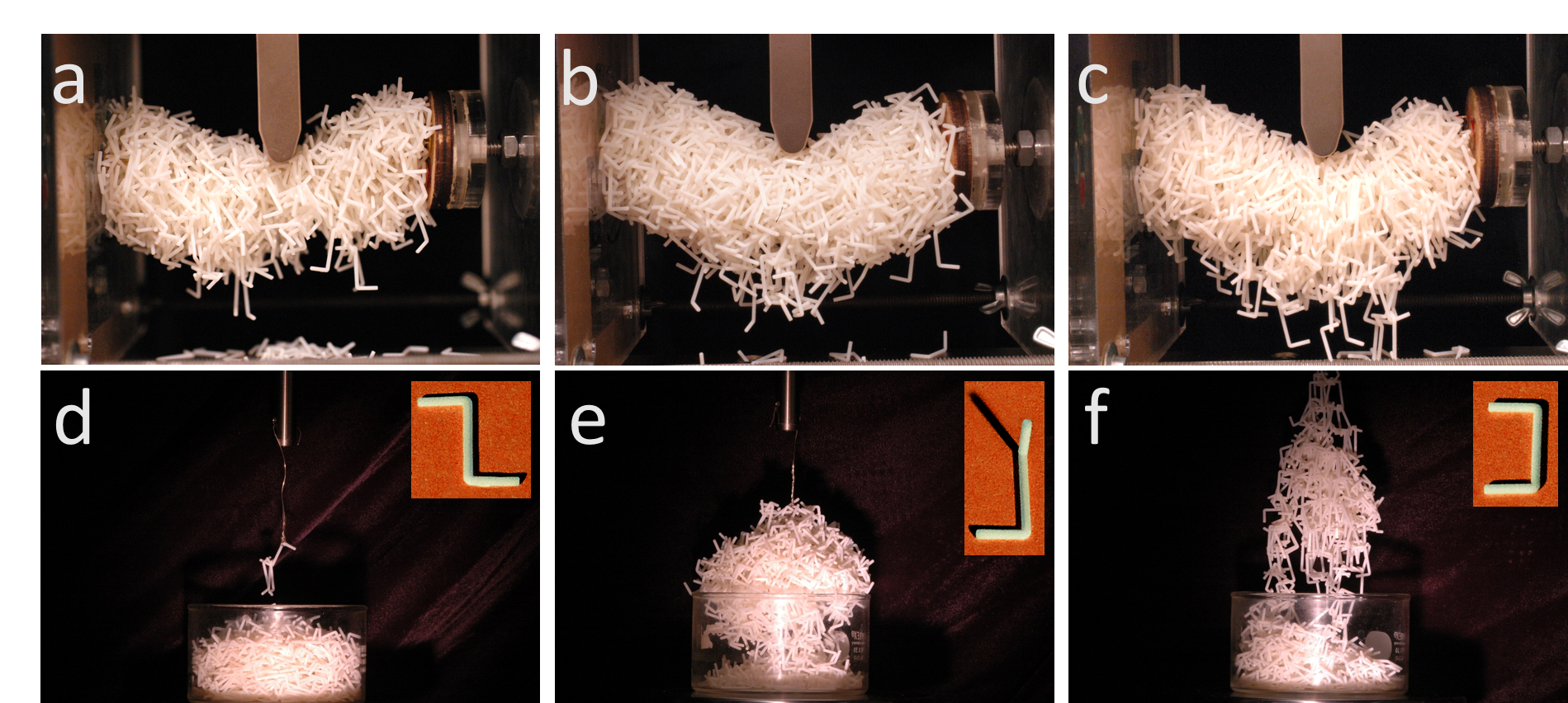}
\caption{Pulling and bending tests for Z (\textbf{a,d}), $Z_{90}$ (\textbf{b,e}) and U (\textbf{c,f}) aggregates. \textbf{a-c}  Final stages of 3-point bending tests (vertical downward extension of 25.0mm) for $\varepsilon_0=0.06$ \textbf{d-f} Particles were poured into a shallow basin and before the last particle-thick layer was poured, we inserted a ``hook'' particle attached to a wire.  The largest particle cluster lifted after three trials is shown here}
\end{figure*}

\begin{figure}
\centering
\includegraphics[width=0.5\textwidth]{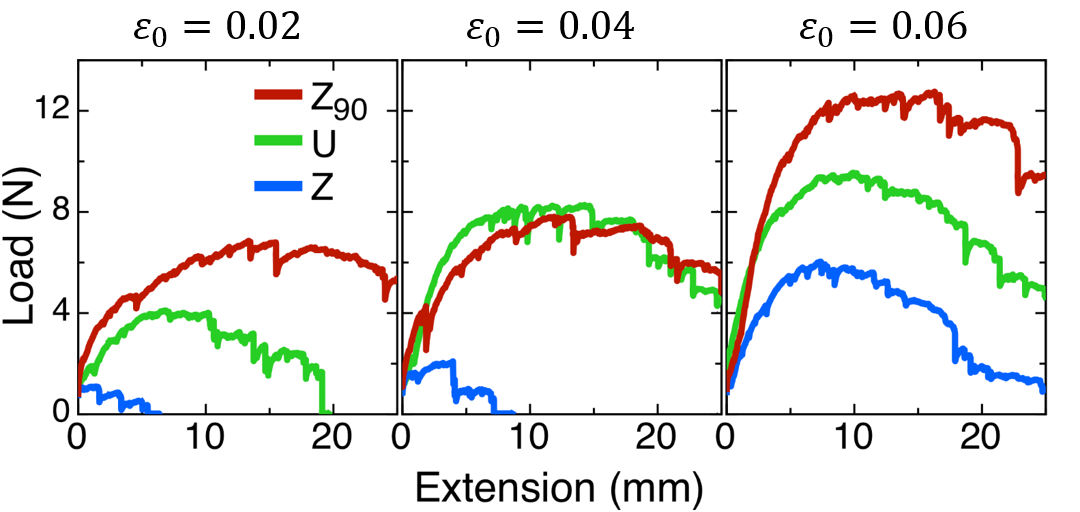}
\caption{3-point bending test results on horizontally oriented columns: the vertical upward force on the indenter as a function of its downward distance into the packing}
\end{figure}

Prior work has shown that rods with aspect ratio $\gtrsim25$ will form rigid packings without need for confinement \cite{Franklin2012,Philipse1996,Desmond2006,Trepanier2010}.  We were indeed able to make stable 2:1 freestanding columns with the aspect ratio 22 rods and compress them to 20\% strain (although generally with significant column deformation), but the relatively weak stress-strain response was roughly linear for the duration of the test.  In contrast, the stress-strain response measured for the particles of the Z-family, both acrylic and 3D printed, exhibited strong strain-stiffening behavior whereby the stiffness of the packing (i.e., the instantaneous slope of the stress-strain curve) increased as the packing was compressed (Fig. 5).  Among 3D printed particles, the Z particles strain-stiffen the most intensely.  If we fit these average stress-strain curves in the low-strain regime $\varepsilon\leq0.05$ to a quadratic polynomial, we can extract both the initial stiffness and a strain-stiffening coefficient.  The latter is roughly an order of magnitude larger for the Zs compared to U and $Z_{90}$ particles.  

The acrylic Z particles are more rigid than their 3D printed counterparts; accordingly the aggregate is much stiffer under compression.  However, the acrylic particles exhibit lower sliding friction against each other and their arms tend to break more easily under tension than for the more pliant 3D printed versions. This explains the much more pronounced dips in individual stress-strain curves for the acrylic Zs (which persist through averaging in the stress-strain response shown in Fig. 5), which are signatures of local failure events inside the packing followed by subsequent self-healing and recovery. This is also the reason we limited the applied strain to 10\% for the acrylic Zs.  In contrast, the columns comprised of 3D printed particles were able to sustain strains of 20\% without damage.

During uniaxial compression the columns formed from Z, $Z_{90}$ and U particles compress along the vertical axis but do not bulge radially outward enough to fail and crumble. As a result the free volume per particle decreases as strain increases and the particles become more and more constrained.  This behavior suggests that Poisson's ratio for aggregates of these particle types is near zero, a nontrivial quality for granular materials which also goes hand in hand with a small B/G ratio, such that the packing would rather compress axially than shear outwards.  A remarkable consequence is that, as long as the prestrain is maintained, the freestanding columns can be handled and even rotated into a horizontal position without falling apart.

Whereas the uniaxial compression tests measured the stiffness along the axis of compression, the stiffness in the transverse direction is also of critical importance when considering structural stability.  In the 3-point bending experiment, we measured the force response as the horizontal column was indented, which characterized the column's resistance to both compression in the portion of the column facing the indenter and dilation on the opposite side (Fig. 6a-c).  Note that the general structure of each indentation-force curve shown in Fig. 7 was the same: an initial linear response, a peak in the force, and a slow decay with further indentation as particles were lost from the dilating side and the column deteriorated.  Interestingly, the eventual failure during bending came from loss of contact between the ends of the columns and the confining endcaps (note the ``peeling'' away of the edges of the column from the endcaps in Fig. 6a-c) suggesting global deformation under bending not unlike that of an ordinary beam made out of a continuous material.  In tests with Z and U aggregates where the pre-strain was low, this slow disintegration of the column typically led to failure (seen in Fig. 7 as a drop in the load to zero) before the indentation reached 25mm, half the column's diameter and our preset end of the test.  As $\varepsilon_0$ increased for a given particle shape, we saw higher peak loads and failure points at larger extensions, if failure happened at all during the experiment.  For a given pre-strain, the shapes were also generally distinguishable based on these differences, with the $Z_{90}$ aggregates the most rigid, followed by U and then Z.  

How easily the different particle types can entangle, and therefore cope with the tensile forces during dilation, can be demonstrated directly by a simple pull test as shown in Fig. 6d-f. In this test, an individual particle is pulled out of an as-poured aggregate by the Instron, and the images show the size of the largest clump of entangled particles that stayed attached after three trials. The clump size qualitatively predicts the trend we see in the bending test: particles which dissociate more easily in the as-poured state, which we equate to flowability and the ease with which particles can shear past each other, produce aggregates that are softer in the transverse direction for low $\varepsilon_0$.

The vibration test demonstrated that higher pre-strain on a column also increases its resistance to dissociation by vibration (Fig. 8).  For the smallest pre-strains ($\varepsilon_0\leq0.01$) the three shapes' behavior was easily distinguishable and showed the same particle shape ordering of rigidity as with the 3-point bending test.  This trend continued until $\varepsilon_0=0.02$, beyond which the three shapes' failure points in the shaker were indistinguishable.  The Z aggregates strengthened the most with initial pre-strain, which matches their having the highest strain-stiffening coefficient.

\begin{figure}
\centering
\includegraphics[width=0.48\textwidth]{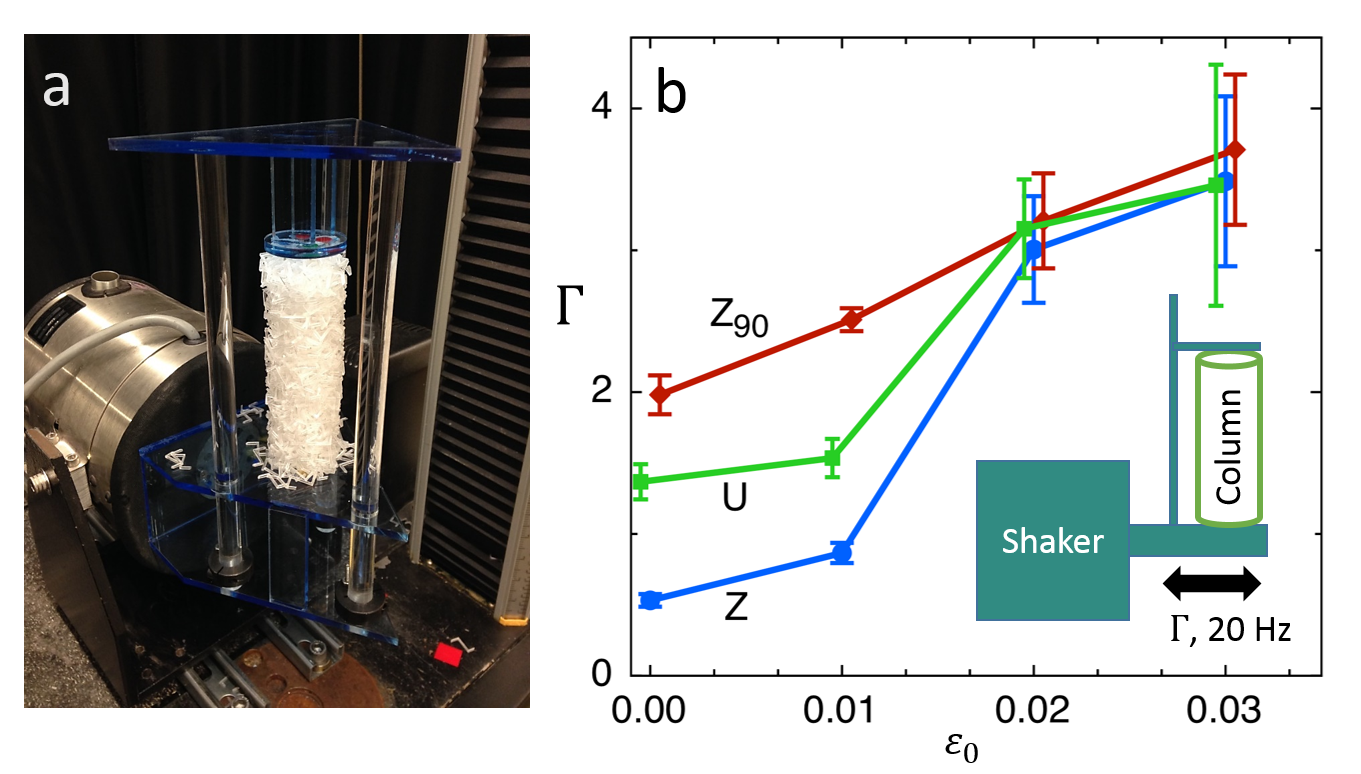}
\caption{\textbf{a} Experimental setup for the vibration test, with a column of Z particles in place.  \textbf{b} Peak acceleration $\Gamma$ at which failure sets in, as a function of pre-strain $\varepsilon_0$.  Data points are offset by $\pm5\times10^{-4}$ in pre-strain for visual clarity. Inset: Sketch of the experiment in panel \textbf{a}, showing the column clamped at a fixed pre-strain and shaken horizontally at 20Hz until failure}
\end{figure}

One aspect that becomes apparent when working with these particles is that they tend to lie flat when poured.  Using particle position data from simulations, we define the polar angle $\theta$ as the angle the backbone makes with respect to the vertical axis, and clearly see that all of the particles studied showed a preference to have their longest section lie closer to $\theta=\pi/2$ (Fig. 9).  For the two planar shapes, Z and U, we can go one step further and measure the tendency of the two legs to also lay flat by defining a vector perpendicular to the plane in which the Zs and Us lie, and tracking the distribution of the angle $\Omega$ between that vector and the vertical; both Zs and Us show the same preference for laying horizontally (Fig. 9, inset).  In Fig. 9 all data shown are from as-poured packings, though changes due to compression and/or confinement removal were found to be insignificant.

\begin{figure}
\centering
\includegraphics[width=0.9\linewidth]{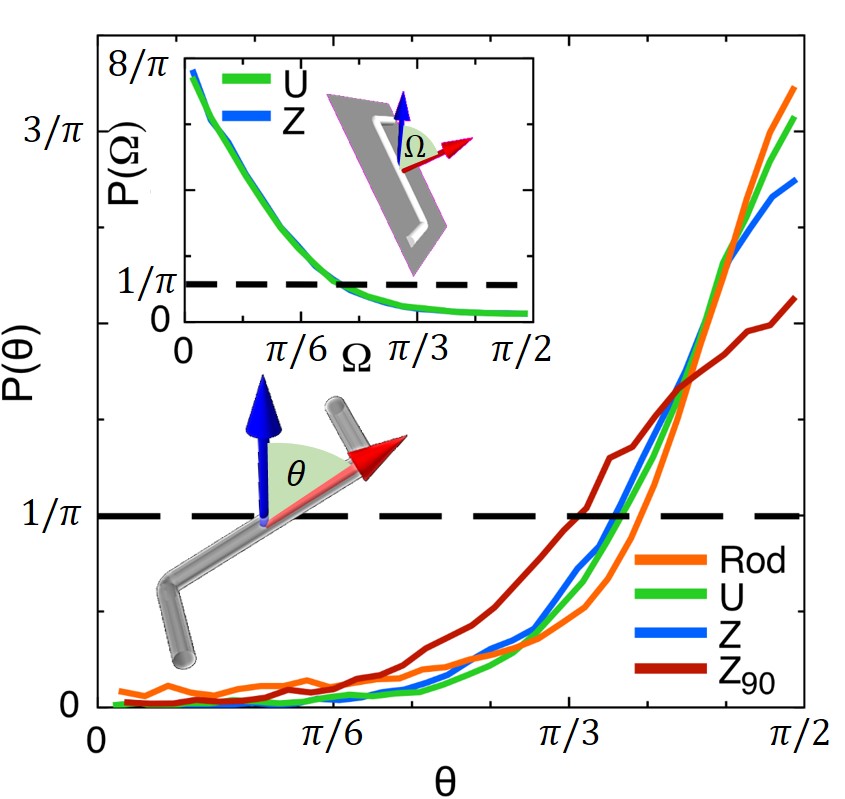}
\caption{Distribution of the polar angle $\theta$ between the vertical (gravity) axis and the longest side (the backbone) for each of the shapes in the packing.  The black dotted line represents total isotropy.  Due to symmetry, only half the range of $\theta$ is displayed.  Inset: Angular orientation of the plane of Z and U particles. $P(\Omega)$ is the distribution of planes with normal vector at angle $\Omega$ with respect to the vertical}
\end{figure}

\begin{figure*}
\centering
\includegraphics[width=1.0\linewidth]{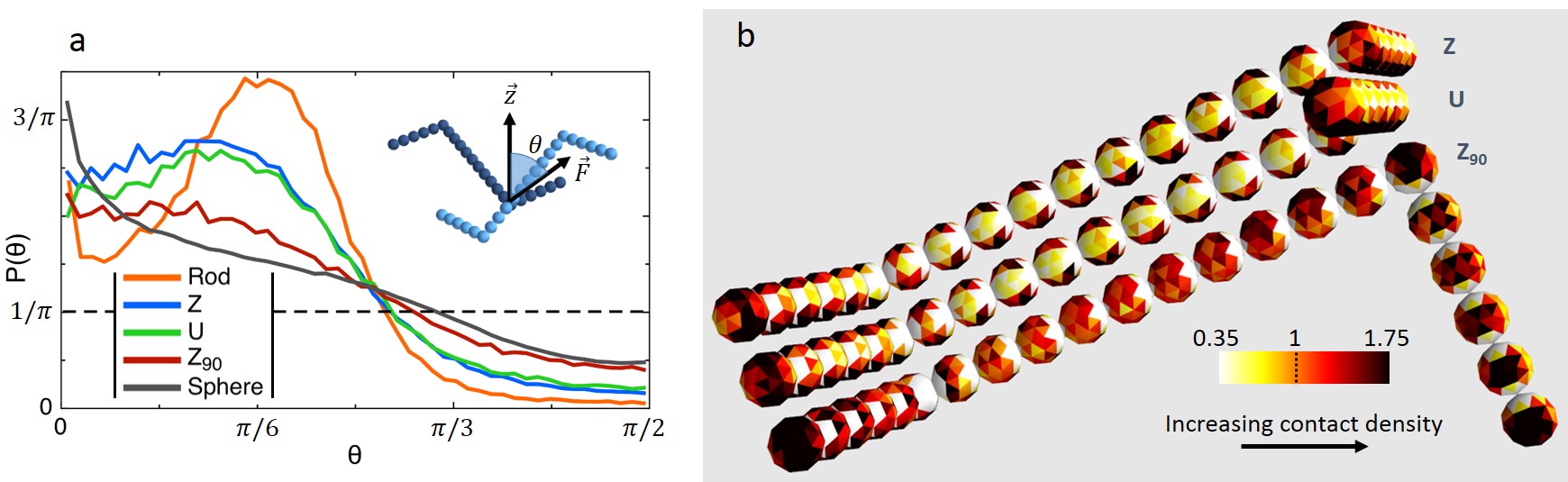}
\caption{\textbf{a} Distribution $P(\theta)$ of contact force directions as a function of polar angle $\theta$ with respect to the vertical. Data are for pre-strain $\varepsilon_0=0.1$.  Note this includes both normal and tangential components.  The black dotted line indicates a fully isotropic distribution. \textbf{b} Histograms showing where contacts occur on a particle.  The color mapping equates 1 with equal uniform probability of a contact over the entire surface area of a particle}
\end{figure*}

This ``pancaking'' of particles creates an anisotropy in the packing that focuses the axial stress propagation and minimizes radial stresses which could lead to expansion in the absence of confinement.  Such anisotropic stress propagation was seen previously for elongated particles in two dimensions \cite{Hidalgo2009,Hidalgo2010}, and is corroborated by the fact that contact forces are predominantly vertical (Fig. 10a).  Note the preference for $\theta=\pi/6$, which we take to be a combination of predominantly vertical stress propagation and an artifact from the way we represent these particles out of bonded spheres.  It could arise from horizontal particles contacting where one particle's sphere is wedged between two spheres from the other in an equilateral triangle configuration; note, however, that if this set of contacts were not oriented vertically, the polar angle $\theta$ would be greater than $\pi/6$.  

The contact locations on a single particle are also far from isotropic: for the planar particles (Z, U), contacts are more often found pressing perpendicular to the plane of the particle, which is seen in Fig. 10b as a higher contact density on the poles of the sub-particle spheres.  These perpendicular contacts pin the prone particles from rotating out of plane and directly limit two of their three rotational degrees of freedom, an effect which intensifies as the axial stress is increased.  The $Z_{90}$ particles cannot be defined by a single plane, which explains the more uniform distribution of contacts we see for them in Fig. 10b.  However, we expect a predominance for one of the two arms to be horizontal, which would then display a similar pinning force anisotropy to what is seen for the Z and U particles.  This leads to the gradient in contact density seen along the backbone of the $Z_{90}$ particle: the in-plane direction on one side of the $Z_{90}$ particle transitions into the perpendicular direction on the other side, and equivalently an equatorial point on a sphere on one side transitions into a pole on the other.  This three-dimensional character means the sprawling arms hinder rotations in a packing even before any compression occurs, which is why the $Z_{90}$ are the most rigid without any pre-strain and why the planar shapes show the largest increase in rigidity as pre-strain is increased (as evidenced by the uniaxial compression (Fig. 5), bending (Fig. 7), and vibration experiments (Fig. 8)).

To examine the stress propagation further, we evaluated the normal force component $F_n$ of each contact in a packing, defined as the force component parallel to the vector connecting the two contacting spheres (which consequently ignores friction).  The distribution of normal force magnitudes, normalized by their mean, $\bar{F_n}$, is $P(F_n/\bar{F_n}) =P(f)$.  As has been established for spheres, the tail of this distribution for large normal forces dies off exponentially \cite{Liu1995,Blair2001,Ohern2001,Erikson2002,Snoeijer2004}.  Hidalgo et al. showed that, at least in two-dimensional packings, elongated particles generate a $P(f)$ that has a more Gaussian character, suggesting a more uniform force transmission through the packing.  What we find, in contrast, is that the highly elongated shapes studied here in 3D packings all had an exponential $P(f)$ tail, showing a higher likelihood of large forces than for a sphere packing (Fig. 11).  This includes straight rods and it seems independent of compression up to the pre-strain levels of 10\% explored in our simulations (Fig. 11 inset). On the other hand, for sphere packings the effect of compression is clear at 10\% pre-strain, producing a Gaussian peak near $f=1$, as seen previously in experiments with packings of rubber spheres \cite{Erikson2002}.  As seen in Fig. 11, the elongated shapes both felt larger normal forces and had a higher proportion of weak forces, defined as $f<1$: as high as 69\% for the rods compared to only 57\% for spheres.  This supports the idea that packings of these elongated shapes are held intact by few, large forces and stabilized by many smaller forces, as seen in anisotropic stress propagation through force chains' strong and weak force networks \cite{Majmudar2005}.  

\begin{figure}
\centering
\includegraphics[width=0.45\textwidth]{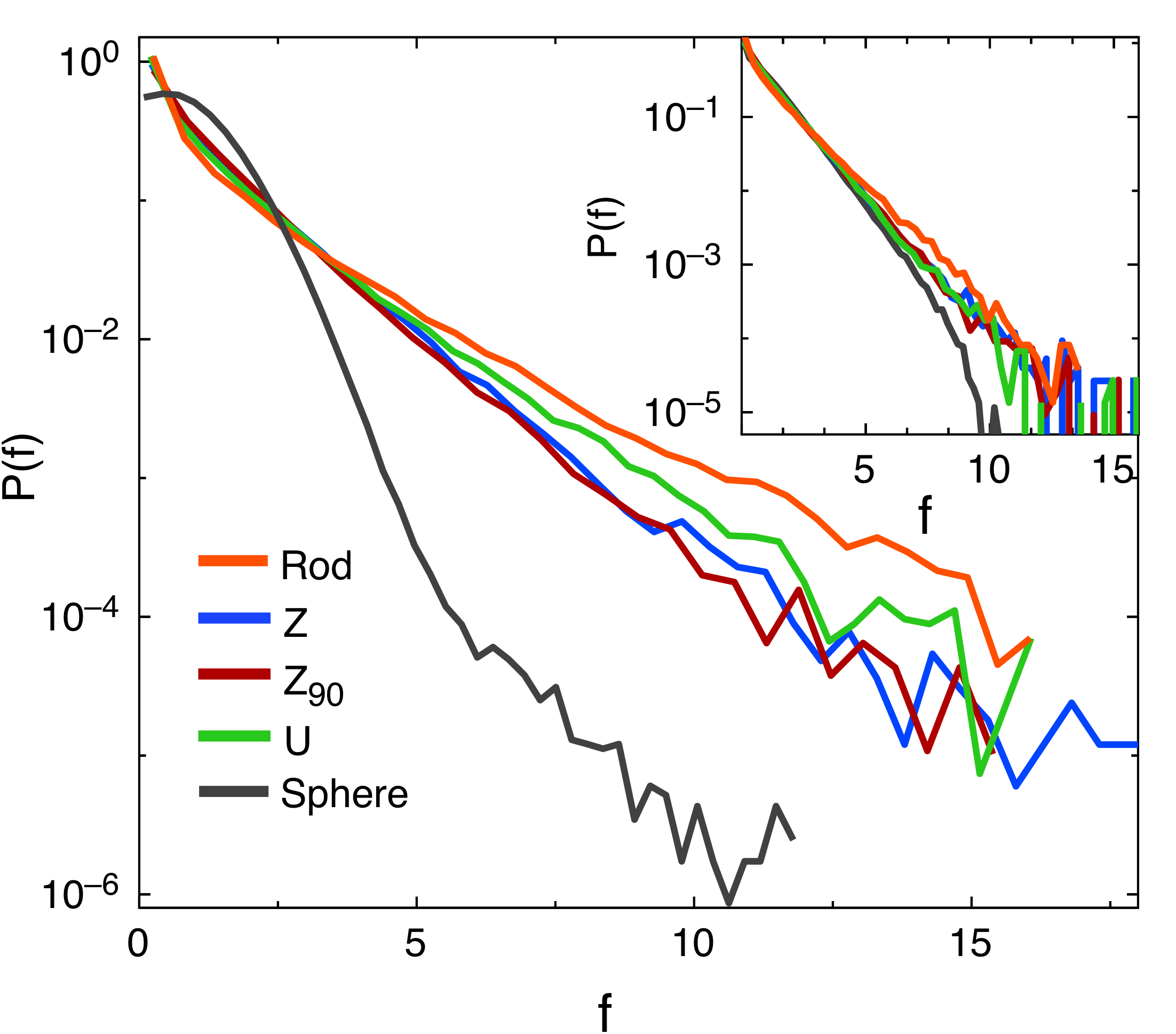}
\caption{$P(f)$ after 10\% pre-strain, averaged over 8 packings for each shape.  Only the sphere packings have radial confinement.  Inset: $P(f)$ for the as-poured state, before any compression is applied}
\end{figure}

The differences seen in these force distributions likely lie in the fact that, as opposed to more convex particles, the elongated, highly anisotropic shapes we studied experience high torques caused by their contact forces.  Particles in a packing where forces are primarily through their centers of mass will reposition until the compressive stresses on the exterior match those in the interior.  On the other hand, particles in a packing with large torques and significant entanglement are easily rotationally frustrated and, as a consequence, can sustain larger contact forces while still maintaining global staticity and equilibrium.

In essence, we see that in these packings the compressive stresses propagate anisotropically, axially down the column, pinning particles and inhibiting outward radial shear in the process.  As such, the strain-stiffening, self confining behavior follows.  Our simulation data showed no significant change in particle orientation as a function of pre-strain for any of the shapes, nor did the average contact number increase any faster with pre-strain than for more convex shapes which do not exhibit self-confining behavior.  The orientational and force anisotropies seen in simulations were actually more pronounced for the rods than for the Z-family particles, yet in experiment the rods were far weaker under compression and showed no strain-stiffening.  This suggests that the perpendicular arms also play a critical role in the rigidity of the Z-family packings; the rod's axial rotational degree of freedom cannot easily be pinned by the vertical contacts as it can for the Z-family particles, due to their extended arms.

\section{Concluding Remarks}

We were interested in finding a granular material, defined by the particle shape and nothing else, that could be used as a building material for architectural structures such as arches and slender columns.  This required it to be controllably switched between two different states: one which flowed easily enough to be poured into any general space, and another which jammed rigidly enough to be solid without radial confinement and under changing external stresses.  We identified a promising family of shapes, represented by the Z, U, and $Z_{90}$.  Not only were packings of these shapes rigid enough under just their own radial self-confinement to support additional axial load, but this rigidity was found to increase (and therefore be tunable) by axial pre-strain.  Furthermore, within this family of shapes, behavior can be tuned by a single parameter $\gamma$: $Z_{90}$ particles form a rigid packing in the as-poured form without any pre-strain, U particles form chains that strongly resist dissociation, and Z particles flow most easily when unconfined but show the largest increase in rigidity as pre-strain is applied.

Through simulating random aggregates of these shapes, we found that the unusual properties stem from anisotropy in the as-poured packings; the particles preferentially orient themselves horizontally relative to the direction of gravity, an ordering which gives rise to primarily axial stress propagation.  Both the orientational anisotropy and the force distributions were similar for all elongated shapes tested, including rods, even though behavior in experiment differed greatly; this suggested that the bent arms hinder the sliding of particles past one another and prevent particle rotation, two aspects that are crucial for the self-confinement of these packings.

\begin{acknowledgements}
We would like to thank Leah Roth, Victor Lee, and Sid Nagel for many illuminating discussions. This works was supported by the NSF through CBET-1334426 and by the University of Chicago’s Arete B.I.G. program.  Use of the Chicago MRSEC’s shared experimental facilities, supported by NSF through DMR-1420709, is gratefully acknowledged.
\end{acknowledgements}

\bibliographystyle{spmpsci}      
\bibliography{Master}   

\end{document}